# Quantum Computing and the Future Internet

By Tajdar Jawaid - ISSA member, UK Chapter

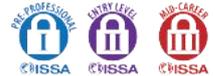

This article discusses quantum computing key concepts, with a special focus on quantum Internet, quantum key distribution, and related challenges.

**Abstract**

One of the biggest concerns among cybersecurity professionals these days is the hype around quantum computing, its incomprehensible power, and its implications. The advancement in quantum computing has the potential to revolutionize our daily lives, but it can also completely break down the Internet as we know it. Mathematicians and physicists have developed algorithms based on quantum computing, which can change the Internet security paradigm. This article discusses quantum computing key concepts, with a special focus on quantum Internet, quantum key distribution, and related challenges.

Classical computing has done an amazing job, revolutionized our daily lives, and has a global impact. From television, the Internet to mobile phones, from booking train and air travels to satellite navigation, everything involves classical computing. In general, we see around us the wonders of classical computing, but we rarely discuss or are aware of the limitations of classical computing such as solving exponential problems. For instance, classical computing has the following limitations or is not very good at finding optimal solutions, to say the least, in the following scenarios:

- **Optimization:** Finding the optimal combination of N numbers of given inputs. One examples is the famous "Traveling Salesman Problem," which is to find the shortest path while traveling between N number of cities. The problem gets quickly exponential with N number of routes of different lengths.
- **Chemistry:** Simulating a molecule on classical computing to find out the reaction rates, synthesis, etc., for instance, simulating molecules like ammonia ($NH_3$) is computationally intractable on classical computing.
- **Factorization:** Finding the prime factor of a large composite number is almost impossible using classical computing, due to the exponential complexity. This limitation of classical computing is the basis of Internet cryptographic solutions, now under threat by quantum computing.

## Quantum mechanics

Quantum mechanics is the field of quantum physics that relates to the study of elementary particle motion, energy, and properties. These elementary or subatomic particles are photons, electrons, the nucleus of an atom, etc. Quantum mechanics describes the interactions, motions, and energies of these sub-atomic particles mathematically and reveals key properties of particles at the quantum level. Characteristics like *quantum superposition* and *quantum entanglement* are the foundation of quantum computing. These key concepts of quantum mechanics and characteristics of quantum particles have been developed throughout the 19th and 20th centuries.

However, the application of quantum mechanics in the computing world was first envisioned in 1980 by Russian-German mathematician Yuri Manin. In his paper, "Computable and uncomputable," Manin discusses the limitations of classical computing and the exponential power of quantum computing due to the nature of superposition and entanglement principles of the quantum world [20]. In 1981 Richard Feynman refined the idea and showed that it is impossible to simulate the quantum system with the use of classical computing [7]. Since the 1980s there have been breakthroughs and developments that have shifted quantum computing from the theoretical realm to the physical world.

Various companies are developing the first wave of experimental quantum computers. Figure 1 shows the IBM Q [13] and the Google quantum computer [15]. IBM has even made quantum computing available to the public in the cloud for free hands-on experience through IBMQ System One.

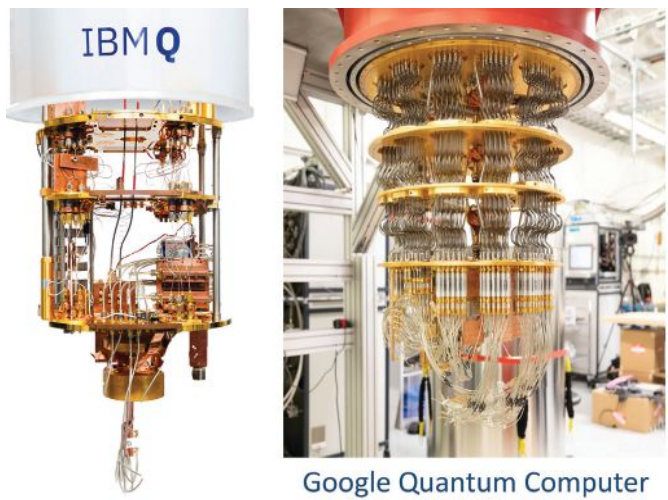

**Figure 1 – IBM Q and Google Quantum Computers**

## Quantum computing

See sidebar "Quantum Mechanics Pioneers" for original papers describing quantum mechanics.

**Quantum superposition:** Superposition is based on Heisenberg's uncertainty principle, which states that the position and momentum of a particle cannot be correctly measured simultaneously. This is due to the characteristic of a quantum particle to be in any or all-states before it is measured.

**Quantum entanglement:** Quantum entanglement states that if any pair of particles generated in a way that they interact with each other, their properties (such as position, spin, momentum, etc.) get appropriately correlated or entangled. They will have exactly opposite but correlated quantum properties in the entanglement state. Once they are entangled no matter how far they are from each other, this correlation remains intact.

**Quantum bits (qubits):** Quantum computers work on quantum bits or qubits. Qubits work on the state of elementary particles like electron, neutron, photons, etc. These elementary particles have internal angular momentum or spins. The states 1 and 0 describe the particle's spin—clockwise/anti-clockwise or up/down—at the point of measurement when there is no surrounding noise. Unlike classical bits, which can have a definitive state of 0 or 1, a qubit can exist anywhere between 0 and 1 simultaneously due to the superposition principle.

For instance, two classical bits can have four possible state combinations (i.e., 00,01,10,11). But at any given state it will only have two bits of information, and to access all four possible states of combinations, the classical computer has to do at least four operations. Whereas two qubits can exist in all four possible states at once, it will require a single operation for quantum computers to access all four states of information stored in two qubits. This quantum advantage increases exponentially with additional qubits. For instance, 4-qubit quantum computers can process 16 bits of information in a single operation and so on. It is thus possible in theory that a 2500 qubits quantum computer can process information that is more than the total number of particles in the observable universe.

## Application of quantum computing

Quantum computing is arguably the most exciting technology of this century, and we are just starting to scratch the surface of it. Quantum computing has the potential to revolutionize the field of medicine by simulating molecules to build new drugs that may help find the cure for diseases and viruses more quickly.

Physicist and mathematicians can solve difficult physical and mathematical problems that have puzzled them for decades. It can help businesses and financial industries to make better forecasts, increase profitability, and achieve sustainable growth. The optimization problems can be solved by simultaneously sampling the whole dataset through the power of quantum superposition to find the optimal results. Quantum computing can also help in adjacent fields like logistics, operations, research, decision making, pattern recognition, machine learning, and quantum simulation of complex material properties and molecular functions.

However, from the infosec point of view, the real threat is breaking of the hardness of factorization. A quantum computer can factor numbers exponentially faster than classical computers as shown by Shor's algorithm [17]. The Internet is currently dependent on public-key cryptographic schemes such as RSA signatures (RSA), the Diffie-Hellman key agreement protocol, elliptic curve digital signature algorithms (ECDSA), etc. These cryptographic schemes are based on the hardness of integer factorization or the discrete logarithm problem. Shor's algorithm shows that a quantum computer with 4099 perfectly stable qubits can break RSA 2048 encryption in just 100 seconds, which would otherwise take billions of years through classical computing.

## Quantum Mechanics Pioneers

**Niels Bohr:** "On the Constitution of Atoms and Molecules" (1913) and "The Structure of the Atom" (1922)

**Werner Heisenberg:** "The Development of Quantum Mechanics" (1933)

**Einstein, Podolsky, and Rosen:** "Can Quantum-Mechanical Description of Physical Reality Be Considered Complete?" (1935)

**Erwin Schrödinger:** "The Present Situation in Quantum Machanics: A Translation of Schrödinger's 'Cat Paradox' Paper" (1935)

## Post-quantum Internet and cryptography

To secure the Internet in the quantum world, quantum key distribution (QKD) algorithms and communication standards need to be developed. To address the threat posed by the advancement in quantum computing, American and European governments have initiated dedicated projects and workshops. The US National Institute of Standards (NIST) has started the "Post-Quantum Crypto Project" [14] to develop a quantum-resistant suite of protocols. Similarly, the European Telecommunication Standards Institute (ETSI), which is one of the three European standards organizations, has started "ETSI-IQC Workshops on Quantum-Safe Cryptography" [6]. Both these standards organizations are pulling scientists together around the globe to address and develop a suite of cryptographic protocols and standards for a quantum Internet.

### Quantum key distribution basics

The first quantum key distribution protocol was developed by Charles Benett and Gilles Brassard in 1984, also known as BB84. The BB84 protocol was developed on the basis of the Heisenberg uncertainty principle. The QKD protocols so far developed fall in the following two categories:

- Protocols based on the Heisenberg uncertainty principle (e.g., BB84) [2]
- Protocols based on quantum entanglement (e.g., E91) [5]

### BB84 Heisenberg uncertainty principle-based Protocol

The BB84 protocol shows particles such as polarized photons can be used to develop and distribute cryptographic keys. Figure 2 describes how bits can be encoded with the polarization of the photon state in BB84. The binary 1 can be encoded using 90° or 135° photon polarization and similarly the binary 0 can be encoded using 45° or 0° photon polarization. The base 1 uses a rectilinear basis for photon polarization, while base 2 uses a diagonal basis for photon polarization for encoding. Figure 2 and 3 illustrated BB84 visually.

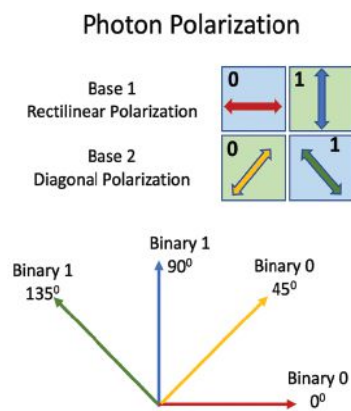

**Figure 2 – Photon polarization in BB84**

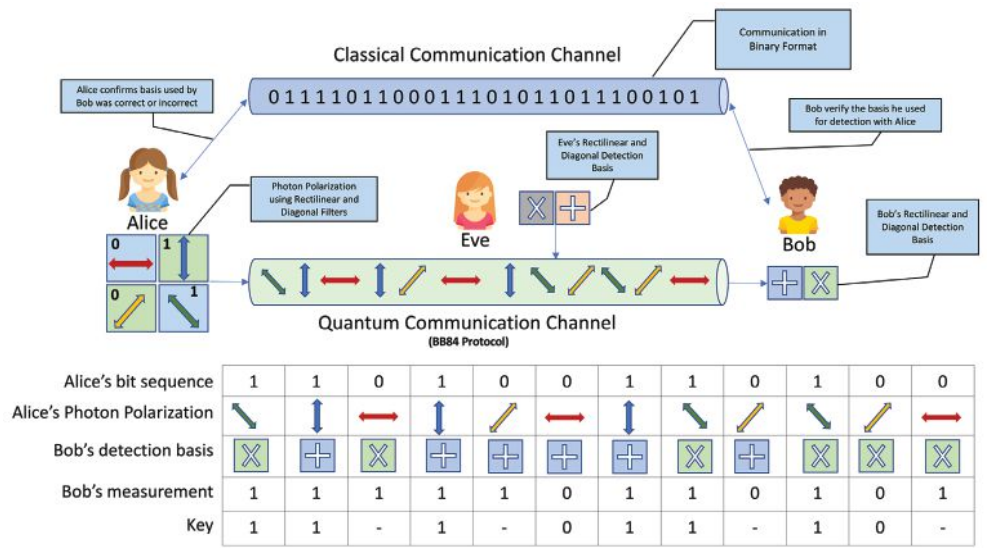

**Figure 3 –BB84 quantum key distribution**

In Figure 3 Alice and Bob decided to establish a secure communication channel using both quantum and classical communication channels. To start the process Alice will use a photon polarization filter to encode binary 1 and 0 using photons polarization states. Alice will send a random bit sequence of 0 or 1 using both polarization bases randomly.

Bob will use a rectilinear or diagonal basis to decode the values without knowing which sequence of basis Alice used to encode the data. Through this Bob will create two tables of bits, one for each basis (i.e., rectilinear and diagonal). Due to the random mix of bits, there is a risk eavesdropper Eve could intercept Alice's bits, decode and resend the bits back to Bob. This will create roughly 50 percent disagreement between Alice and Bob where they think their sequence of the basis used for encoding and decoding should coincide. This is due to the fact that Eve will destroy the information as soon as she reads it to decode. Bob's information will further be reduced due to the photons lost in transmission due to noisy channels.

Once the quantum communication phase has ended, Alice and Bob will use the classical channel to verify the communication through a process called *sifting*. They will identify which protons were received successfully, and the sequence of the bases used for encoding and decoding. If the quantum communication did not have any interference, Alice and Bob agree on the bits encoded/decoded using the sequence of the bases used.

In the case of eavesdropping, the encoding and decoding will produce different results on the sequence where they think their results should match. They will discard all those bits where they use the same sequence but have different results. Where comparison agrees they will use those bits as a one-time pad to send communication over the classical channel. Once this one-time pad is used up, this protocol is repeated to generate a new one-time pad over the quantum channel for subsequent information.

In 1992 Charles Bennett proposed a simplified version of BB84 referred to as BB92. BB92 describes that it is not nec-essary to use four polarization states: the 0 and 1 can be encoded using just two polarization states. 0 can be encoded at 0° rectilinear bases, whereas 1 can be encoded at 45° diagonal bases without compromising the security [3].

### E91 entanglement-based Protocol

In 1991 Artur Ekert reported the application of Bell's theorem in the quantum key distribution process. The basic premise of Ekert's paper [5] is the fact that entanglement is inherently secure. The entangled qubit properties cannot be copied or reproduced within the known laws of nature. Hence, this property should be applied to quantum key distribution. Applying the Alice and Bobs example to the E91 protocol, Alice and Bob both will receive the entangled photons from the same source. These entangled photons are the exact opposite of each other due to the entanglement principle discussed above. Alice and Bob can agree on a key basis as in BB84 and can invert their keys to reveal the secret key.

### QKD security concerns

QKD protocols promise the highest possible level of security due to the known laws of quantum physics that cannot be violated. However, QKD protocols are still susceptible to attacks due to imperfect communication channels.

**Man-in-the-middle attack** (eavesdropping) or due to **noisy imperfect transmission channels**. It is not easy to differentiate the two errors; hence, both are assumed to be eavesdropping. To mitigate these errors there are two techniques (Information Reconciliation and Privacy Amplification) proposed in 1992 by Bennett and Brassard with their colleagues [4].

1. *Information reconciliation* is the process to remove the errors caused by noisy channels.

2. *Privacy amplification* is the process of reducing or eliminating Eve's partial information about Alice and Bob's key.

**Photon number splitting attack** is another kind of attack in addition the a man-in-the-middle attack. Due to noise in the quantum channel, it is impractical to produce and detect a single photon. In practice a laser pulse is used to produce a small number of photons. In the case of Alice and Bob transmitting the photons through a laser, Eve can split off the single or small number of photons to measure, allowing the rest of the photons to still be transmitted to Bob. This will help Eve have the secret key information without disturbing the whole communication and, hence, without being noticed. To mitigate this, a solution has been developed by Lo et al, using decoy signals randomly in a way that Eve cannot differentiate between the decoy and non-decoy photon pulses [12].

### Quantum communication protocol for quantum Internet

By using quantum entanglement capabilities scientist are researching to develop quantum communication for the future Internet. As we know, entangled particles can have information regardless of how far apart they are. However, the entanglement itself is very fragile; it doesn't hold after a certain distance. As particles cannot be copied (no cloning theorem) and cannot maintain the status for long, they quickly lose their entangled status [19]. Scientists are working on quantum repeaters that will be used to extend the transmission of entangled particles over large distances through a technique called entanglement swapping to achieve quantum teleportation [1].

## Quantum computing challenges

As quantum computing is still in its infancy, there are many problems that have to be overcome. The following are some of the key challenges faced by quantum computing.

### Quality of qubits

The quality of qubits refers to error correction rates and stability of quantum properties or states held by the qubits, also known as their coherence time. Figure 4 shows the coherence time visually in which a qubit holds its state information. Generally, qubit coherence time is between 50 to 90 milliseconds in trapped-ion systems.

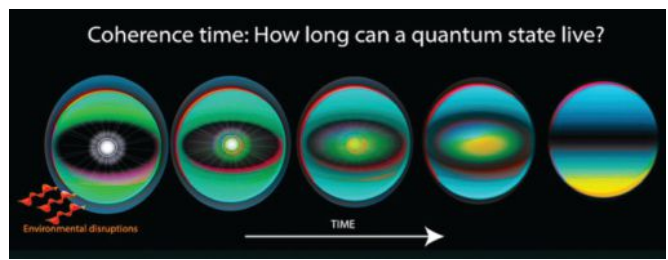

**Figure 4 – Coherence Time** [11]

There are research and experiments that show that environments can be created where the state of qubits endures from 10 to 39 minutes [18] [16]. This opens up new research opportunities for developing this technology further.

### Quantum error correction (QEC)

Computation is required to have error detection and error correction mechanisms and algorithms in place to do the precise calculations. Physicists have developed some very efficient techniques to do large-scale quantum error correction on a two-dimensional qubit grid with about one percent error tolerance [8].

This means that to break classical 2048-bit RSA encryption a total of 220 million physical qubits are required, and it would take around 26.7 hours to break the encryption by finding the factors [8]. This was further refined by Craig Gidney and Martin Ekera in 2019 when they demonstrated that it would require eight hours with 20 million noisy qubits to break 2048 RSA encryption [9].

### Cryogenic cooling

Quantum particles vibrate and fluctuate continuously due to Heisenberg's uncertainty principles. This vibration of quantum particles or atoms is dependent on their mass: the lighter the atom, the more it vibrates. To work with vibrating quantum particles, a state of near zero-point motion energy needs to be achieved. Supercooling to very low temperatures is required to propel the quantum particles to the state of extremely low energy levels so that they can be easily controlled.

Figure 5 visualizes atoms at room temperature and at near absolute zero. The cryogenic temperature ranges from -150˚C (-238˚F) to absolute zero (-273˚C or -460˚F). IBM's quantum computer, for instance, uses 15millikelvin to cool down its quantum computer, which is colder than the temperature of outer space [10]. Working with these low temperatures is not very easy and could be an obstacle to produce commercial quantum computers. But there are breakthroughs in molecule-sized and atom-sized transistors that once matured enough may help overcome this challenge.

### No-cloning theorem

A quantum state cannot be copied—no cloning theorem—due to the conservation of quantum information law [19].

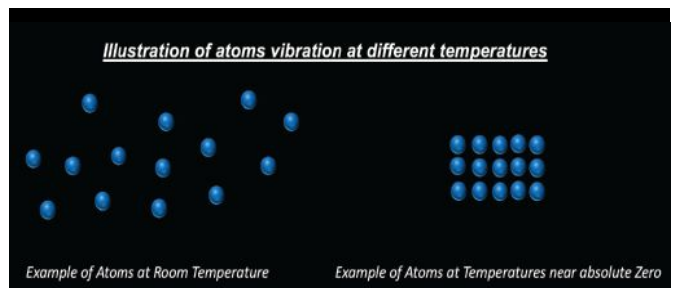

**Figure 5 – Atom vibrations at different temperatures**

This represents a challenge while transmitting, storing, and processing the information through quantum computing without the ability of copying and duplicating the qubits. Special algorithms, techniques, and technologies need to be developed to overcome this challenge.

## Conclusion

Quantum computing is a promising and emerging field that has great prospects to revolutionize our daily lives. But there are challenges in quantum computing, though these challenges can be categorized as a "teething" problem instead of limitations. There are recent breakthroughs and advancements in technology, new research, and development in this field that will make the quantum revolution possible in the foreseeable future. Unlike the hype, it is safe to assume that there is no immediate threat to the current public-key infrastructure through quantum computers, as it is still requires a quantum computer with millions of qubits to break current cryptography. The parallel research and advancement in the field of quantum cryptography algorithms like NIST "Post-Quantum Crypto Project" and ETSI "Workshops on Quantum-Safe Cryptography" will help develop a secure quantum Internet. The practical impact of this technology on our daily lives may be a few decades away, but there is no doubt that once this technology is matured enough, the future of humanity will be very different.

### About the Author

*Tajdar Jawaid, PMP, MS Cybersecurity from University of Dallas, TX, is a security architect working for Telefonica UK. He has around 18 years of experience in security architecture and design. He may be reached at tjawaid@udallas.edu.*

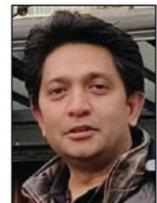